\documentclass[12pt]{iopart}
\pdfoutput=1
\expandafter\let\csname equation*\endcsname\relax
  \expandafter\let\csname endequation*\endcsname\relax
\usepackage[numbers,sort&compress]{natbib} 
\expandafter\let\csname equation*\endcsname\relax
\expandafter\let\csname endequation*\endcsname\relax
\usepackage{amsmath,amssymb,mathtools}

\usepackage{graphicx,color}
\usepackage{enumerate}
\usepackage{subfigure}

\newcommand\be{\begin{equation}} 
\newcommand\ee{\end{equation}}

\begin{document}

\title{Reputation-Driven Voting Dynamics}

\author{D.~Bhat and S.~Redner}
\address{Santa Fe Institute, 1399 Hyde Park Road, Santa Fe, NM, 87501}

\date{\today}
\begin{abstract}

  We introduce the reputational voter model (RVM) to account for the
  time-varying abilities of individuals to influence their neighbors.  To
  understand of the RVM, we first discuss the fitness voter model (FVM), in
  which each voter has a fixed and distinct fitness.  In a voting event where
  voter $i$ is fitter than voter $j$, only $j$ changes opinion.  We show that
  the dynamics of the FVM and the voter model are identical.  We next discuss
  the adaptive voter model (AVM), in which the influencing voter in a voting
  event increases its fitness by a fixed amount.  The dynamics of the AVM is
  non-stationary and slowly crosses over to that of FVM because of the
  gradual broadening of the fitness distribution of the population.  Finally,
  we treat the RVM, in which the voter $i$ is endowed with a reputational
  rank $r_i$ that ranges from 1 (highest rank) to $N$ (lowest), where $N$ is
  the population size.  In a voting event in which voter $i$ outranks $j$,
  only the opinion of $j$ changes.  Concomitantly, the rank of $i$ increases,
  while that of $j$ does not change.  The rank distribution remains uniform
  on the integers $1,2,3,\ldots,N$, leading to stationary dynamics.  For
  equal number of voters in the two voting states with these two
  subpopulations having the same average rand, the time to reach consensus in
  the mean-field limit scales as $\exp(\sqrt{N})$.  This long consensus time
  arises because the average rank of the minority population is typically
  higher than that of the majority.  Thus whenever consensus is approached,
  this highly ranked minority tends to drive the population away from
  consensus.

\end{abstract}


\section{Introduction}

The way people form an opinion about a given issue, such as making a
political decision of choosing a product is a complex social phenomenon.  An
individual's opinion can be influenced by economic factors, advertising, mass
media, as well the opinions of others.  When opinion changes occur only
through interactions between individuals, a natural model for this dynamics
is the voter model
(VM)~\cite{clifford1973model,holley1975ergodic,Cox,Liggett,krapivsky1992kinetics,dornic2001critical,RevModPhys.81.591,R01,krapivsky2010kinetic,baronchelli2018emergence}.
In the VM, each individual, or voter, can assume one of two states, denoted
as $+$ and $-$, with one voter at each node of an arbitrary network.  A voter
is selected at random and it adopts the state of a randomly chosen
neighboring voter.  This update is repeated at a fixed rate until a
population of $N$ voters necessarily reaches consensus.  Each voter is
influenced only by its neighbors and has no self confidence in its own
opinion.

The paradigmatic nature of the VM has sparked much research in probability
theory~\cite{clifford1973model,holley1975ergodic,Cox,Liggett} and statistical
physics~\cite{krapivsky1992kinetics,dornic2001critical,RevModPhys.81.591,krapivsky2010kinetic,baronchelli2018emergence,M03}.
Because of its flexibility and utility, the VM has been applied to diverse
problems, such as population genetics~\cite{Blythe},
ecology~\cite{Zillio,Maritan}, and epidemics~\cite{Pinto}, and voting
behavior in elections~\cite{Gracia}.  However, consensus is not the typical
outcome for many decision-making processes.  This fact has motivated a
variety of extensions of the VM to include realistic elements of opinion
formation that can forestall consensus.  Examples include: stochastic
noise~\cite{Manfred,Boris,Adrian}, the influence of multiple
neighbors~\cite{Castellano-q}, self confidence~\cite{Volovik},
heterogeneity~\cite{Masuda}, partisanship~\cite{XSK11,Masuda2}, and multiple
opinion
states~\cite{deffuant2002can,hegselmann2002opinion,ben2003bifurcations}.

An important extension of the VM that is relevant to this work arises when
either the underlying network or the decision-making rule of each voter
changes with
time~\cite{gross2006epidemic,holme2006nonequilibrium,kozma2008consensus,shaw2008fluctuating,durrett2012graph,rogers2013consensus,Woolcock}.
The latter scenario represents an attempt to account for the feature that the
influence of individuals may be time dependent---some individuals may become
more influential and others less so as the opinions of the population evolve.
A natural way to account for this feature is to assign each individual a
fitness that can change with time.  In a single update, the higher-fitness
voter imposes its opinion on its neighbor and correspondingly, the fitness of
the influencer increases by a fixed amount, while the fitness of the
influenced voter does not change.  This adaptive voter model (AVM),
introduced in~\cite{Woolcock}, leads to a consensus time on the complete
graph that appears to scale as $N^{\alpha}$, with $\alpha\approx 1.45$, a
slower approach to consensus compared to the classic VM.  We will
argue, however, this model exhibits a very slow crossover that masks the
asymptotic approach to consensus.

This AVM also provides the motivation for our \emph{reputational voter model}
(RVM) to help understand the role of individual reputation changes on the
consensus dynamics.  In the RVM, each voter is endowed with a unique
integer-valued reputation that ranges from 1, for the voter with the best
reputation, to $N$, for the voter with the worst reputation, in addition to
its voting state.  In an update, two voters in different opinion states are
selected at random and the voter with the higher reputation imposes its
voting state on the voter with the lower reputation.  After this interaction,
only the reputation of the influencer voter rises, in analogy with the AVM.
As we will show, the effect of these reputational changes significantly
hinder the approach to consensus.  When the population initially contains
equal numbers in the two voting states and the average rank of these two
subpopulations are the same, the time to reach consensus scales as
$\exp(\sqrt{N})$.  This slow approach to consensus arises because close to
consensus the average rank of the minority population is typically higher
than that of the majority.  This imbalance tends to drive the population away
from consensus and thereby leads to a long consensus time.

In Sec.~\ref{sec:models}, we define the models under study: (i) the fitness
voter model (FVM), where each voter is assigned a unique and unchanging
fitness value, (ii) the adaptive voter model (AVM)~\cite{Woolcock}, and (iii)
the reputational voter model (RVM).  In Sec.~\ref{sec:fvm}, we will show that
the FVM has the same dynamics as the classic VM.  In
Sec.~\ref{sec:avm}, we will argue that the consensus time scaling as
$N^{\alpha}$, with $\alpha\approx 1.45$ in the AVM~\cite{Woolcock}, is a
finite-time artifact and that the dynamics of the AVM eventually crosses over
to that of the FVM.  In Sec.~\ref{RVM}, we introduce the RVM and discuss the
role of the time-dependent individual reputations on the opinion dynamics.
In Sec.~\ref{conclude}, we give some concluding remarks.

\section{MODELS}
\label{sec:models}

We begin by defining a set of voter models that culminate with the RVM, which
is the focus of this work.  All our models are defined on the complete graph;
this structure is assumed throughout.

\subsection{Classic Voter Model (VM)}
\label{subsec:VM}

We define the classic VM in a form that is convenient for our subsequent
extensions.  In the VM, voters are situated on a complete graph of $N$ nodes,
with one voter per node.  Each voter is initially assigned to one of two
opinion states, $+$ or $-$.  The number of voters in the $+$ and $-$ states
are denoted by $N_+$ and $N_-$.  The opinion update is the following:
\begin{enumerate}
\itemsep -0.5ex
\item[(a)] Pick two random voters in opposite opinion states.
\item[(b)] One of these two voters changes its opinion.
\item[(c)] Repeat steps (a) and (b) until consensus is necessarily reached.
\end{enumerate}
Figuratively, each agent has no self-confidence and merely adopts the state
of one of its neighbors.  After each update, the time is incremented by an
exponential random variable with mean value $\delta t\equiv N/(N_+N_-)$.

There are two basic observables in the VM: the consensus time and the exit
probability.  The consensus time, $T_N(m)$, is the average time for a
population of $N$ voters to reach unanimity when the initial magnetization,
which is the difference in the density of $+$ and $-$ voters, equals $m$.
For the complete graph, the consensus time is (see e.g.,~\cite{R01})
\begin{subequations}
\begin{align}
  T_N(m) =-N\Big\{(1+m)\ln \big[\tfrac{1}{2}(1+m)\big]
  +(1-m)\ln \big[\tfrac{1}{2}(1-m)\big]\Big\}\,.
\end{align}
We are often interested in the zero-magnetization initial condition, in which
case, we write the consensus time as $T_N$.  The main feature of the
consensus time on the complete graph is that it grows linearly with $N$.

The exit probability $E(m)$ is defined as the probability that a population
of $N$ voters with initial magnetization $m$ reaches $+$ consensus.  The form
of the exit probability is especially simple because the average
magnetization is conserved:
\begin{align}
  E(m)= \tfrac{1}{2}(1+m)\,.
  \label{vm-ep}
\end{align}
\end{subequations}
In voter models where the magnetization is not conserved, the exit
probability is a non-linear function of $m$.

\subsection{Fitness Voter Model (FVM)}
\label{subsec:fvm}

In the FVM, each voter is assigned an opinion state as well as a unique and
fixed fitness that is drawn from a uniform distribution in the range
$[0,F_0]$.  A voter with a larger fitness value is regarded as more fit.  The
opinion update is now:
\begin{enumerate}
\itemsep -0.5ex
\item[(a)] Pick two random voters in opposite opinion states.
\item[(b)] The less fit voter changes its opinion.
\item[(c)] Repeat steps (a) and (b) until consensus is necessarily reached.
\end{enumerate}
The time increment for each update is again an exponential random variable
with mean value $\delta t$.  The crucial feature of the FVM is the
\emph{unique} fitness of each voter; the actual fitness values are
immaterial.  We will show below that the dynamics of the FVM is the same as
the VM.

\subsection{Adaptive Voter Model (AVM)}
\label{subsec:avm}

In our version of the AVM, each voter is assigned a unique fitness that is
drawn from the uniform distribution $[0,F_0]$.  The fitness of each voter
also changes as a result of opinion updates.  The opinion update is given by:
\begin{enumerate}
\itemsep -0.5ex
\item[(a)] Pick two random voters in opposite opinion states, with fitnesses
  $f_i$ and $f_j$.
\item[(b)] The less fit voter changes its opinion.
\item[(c)] For the fitter voter $i$, $f_i\to f_i+\delta\! f$.
\item[(d)] Repeat steps (a)--(c) until consensus is necessarily reached.
\end{enumerate}
After each update, the time is incremented by an exponential random variable
with mean value $\delta t$.  As we shall see, the initial fitness range
$F_0$, the fitness increment $\delta\!f$ in each voting event, and $N$ play
important roles in determining the long-time dynamics.

\subsection{Reputational Voter Model (RVM)}
\label{subsec:rvm}

\begin{figure}[ht]
  \centerline{\includegraphics[width=0.425\textwidth]{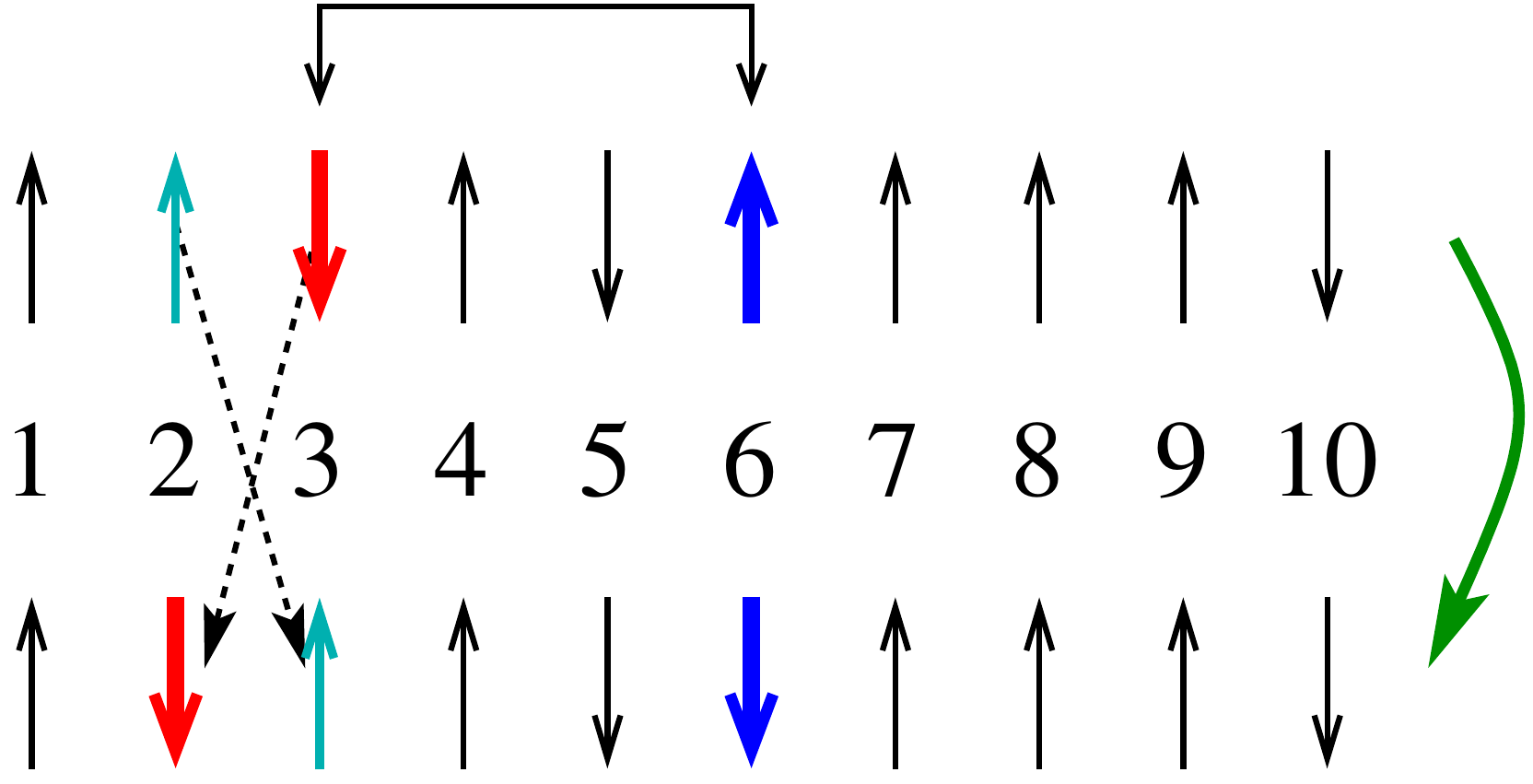}}
  \caption{\small Update event in the RVM.  Voters are arranged in rank
    order.  The voter with rank 3 changes the opinion of the voter with rank
    6. After the voting event, the ranks of the influencer and an adjacently
    ranked voter are shuffled to avoid ties.}
\label{rvm}  
\end{figure}

In the RVM, each voter is assigned a unique and integer-valued reputation, or
rank, between 1 and $N$, with 1 corresponding to the best-ranked voter and
$N$ to the worst-ranked.  The opinion update is given by:
\begin{enumerate}
\itemsep -0.5ex
\item[(a)] Pick two random voters in opposite opinion states, with ranks
  $r_i$ and $r_j$.
\item[(b)] The lower-ranked voter changes its opinion.
\item[(c)] The higher-ranked voter $i$ gains rank, $r_i\to r_i-1$.
\item[(d)] The rank of the voter with rank adjacent to $i$ is adjusted to
  eliminate ties (Fig.~\ref{rvm}).
\item[(e)] Repeat steps (a)--(d) until consensus is necessarily reached.
\end{enumerate}
As we will see, when the population is close to consensus, minority-species
voters are typically well ranked and more likely to influence the majority
rather than be influenced.  This effective bias drives the population back to
equal densities of $+$ and $-$ voters, and leads to a large consensus time.

\section{Dynamics of the Fitness Voter Model (FVM)}
\label{sec:fvm}

The main feature of the FVM is that its dynamics is identical to that of the
VM.  This equivalence will be important to understand the dynamics of the
AVM, that will be treated in the next section.  First
consider the dependence of the exit probability $E(m)$ on the initial
magnetization $m$.  By construction, the fittest voter in the population can
never change its opinion.  Consequently, the final consensus state coincides
with the initial voting state of this fittest voter.  The probability that
the fittest voter is in the $+$ state equals $\frac{1}{2}(1+m)$.  Thus
$E(m)=\frac{1}{2}(1+m)$, as in the VM.

Let us now treat the consensus time.  For the VM on the complete graph, the
initial magnetization uniquely specifies the system.  From this initial
state, there are many trajectories that eventually take the system to
consensus.  To compute fundamental quantities like the exit probability and
the consensus time, we need to average over all stochastic trajectories of
the voting dynamics.  For the FVM on the complete graph, the initial state is
specified by both the magnetization and the fitness of each voter.  The
computation of the exit probability and the consensus time requires
averaging over all stochastic trajectories and over all fitness values.

Thus let us compare the fate of a single pair of voters $ij$ in the state
$+-$ in the VM and in the FVM.  For the VM, this pair changes to either $++$
or $--$ equiprobably.  In the FVM, if the fitness of voter $i$, $f_i>f_j$,
then this pair changes from the state $+-$ to $++$.  However, if $f_i<f_j$,
then this pair changes from $+-$ to $--$.  Since it is equally likely that
$f_i>f_j$ or $f_i<f_j$, then in averaging over all stochastic trajectories
\emph{and} over all fitness assignments, the $ij$ pair in the FVM equally
likely changes to $++$ or to $--$.  Thus the dynamics of the VM, averaged
over all stochastic trajectories, is the same as that of the FVM, when
averaged over all stochastic trajectories and over all initial fitness
assignments.  A detailed microscopic derivation of this equivalence is given
\ref{phi}.

\section{Dynamics of the Adaptive Voter Model (AVM)}
\label{sec:avm}

\subsection{Consensus Time and Exit Probability}
\label{sec:ctep-avm}

\begin{figure}[ht]
\centerline{
    \subfigure[]{\includegraphics[width=0.33\textwidth]{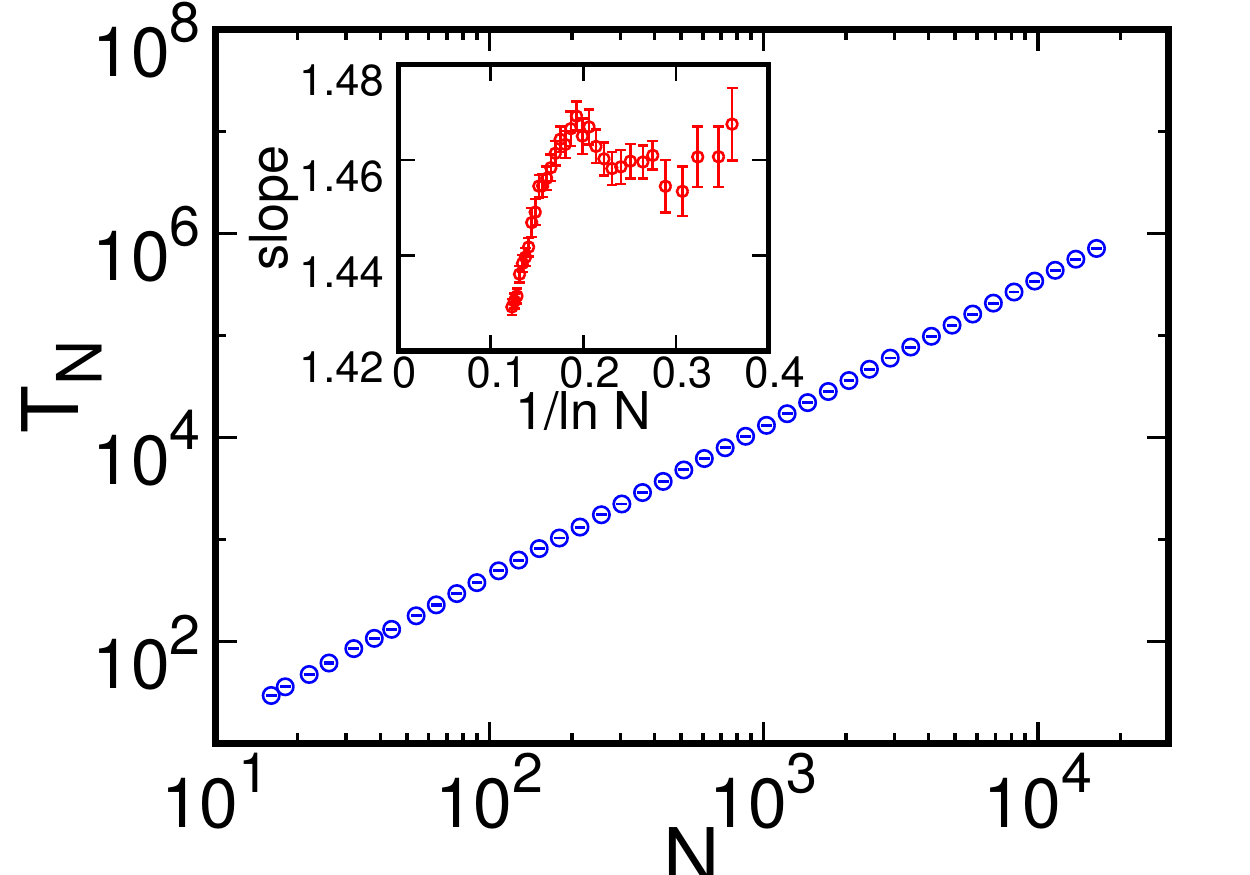}}
    \subfigure[]{\includegraphics[width=0.33\textwidth]{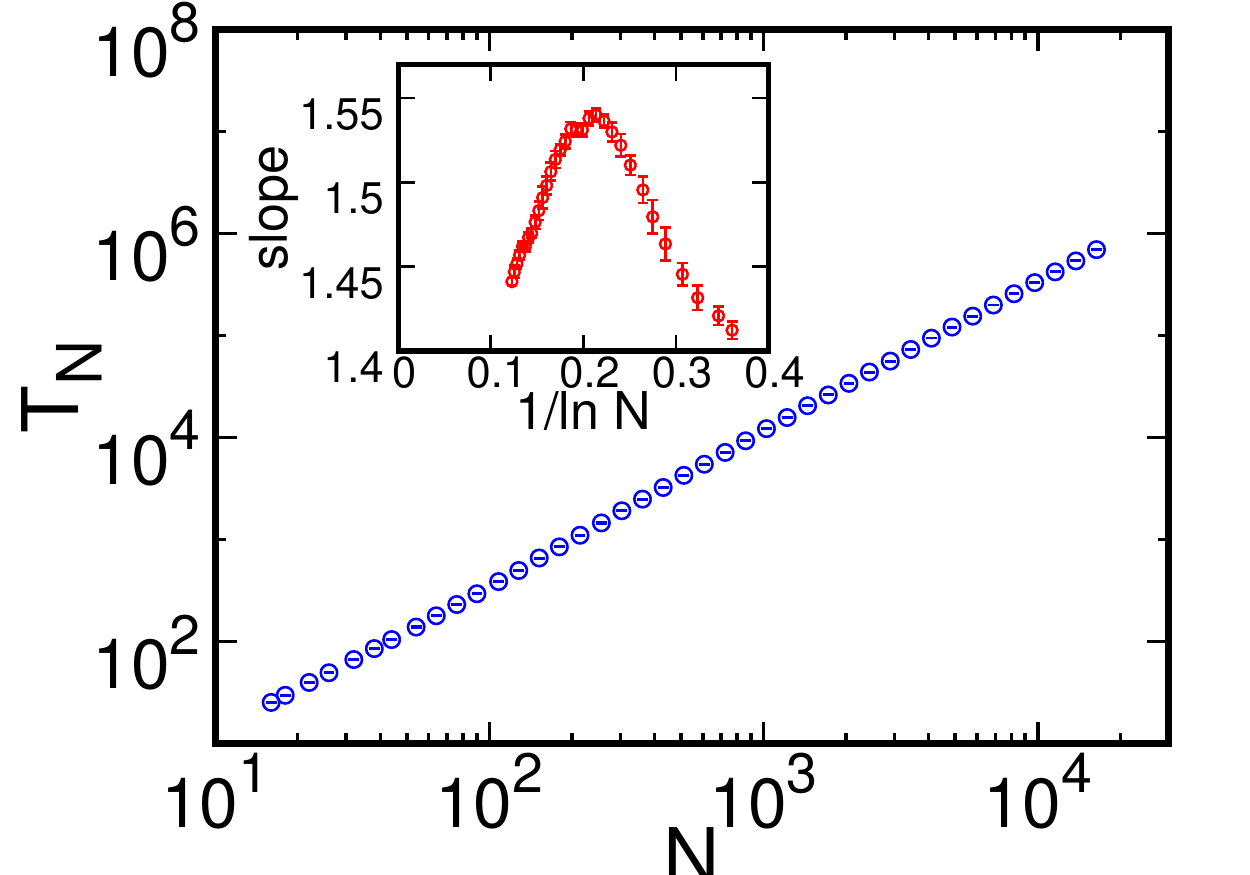}}
    \subfigure[]{\includegraphics[width=0.33\textwidth]{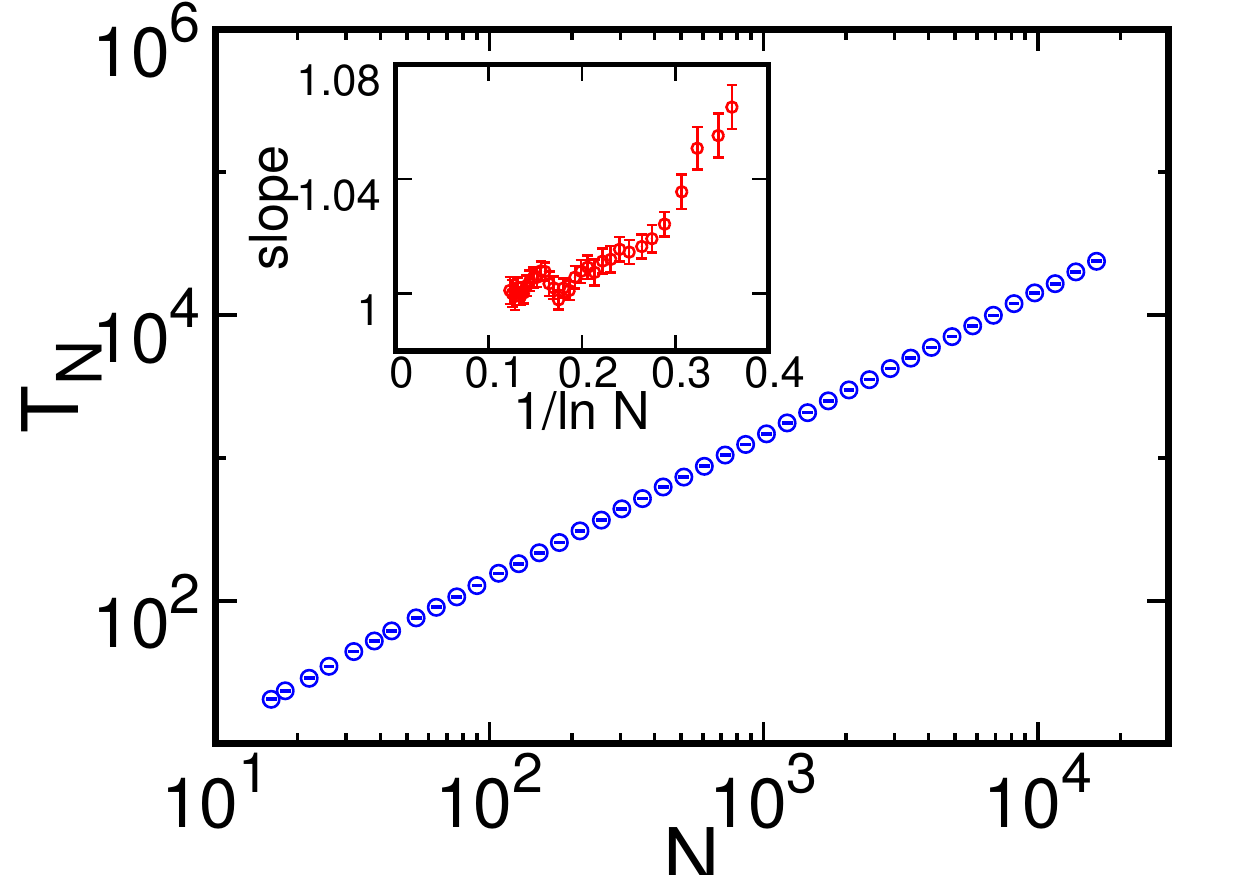}}}
  \caption{Average consensus time $T_N$ versus $N$ for the AVM on the
    complete graph of $N$ sites with: (a) $F_0=1$, (b) $F_0=N$, and (c)
    $F_0=N^{2}$, with $\delta\! f=1$.  The insets show local
    10-point slopes as a function of $1/\ln N$.  The error bars are the
    standard deviation in a linear least-squares fit. }
\label{avm-T}
\end{figure}

In Ref.~\cite{Woolcock}, it was reported that the consensus time scales as
$T_N \sim N^{\alpha}$, with $\alpha\approx 1.45$.  Instead, we will argue
that this exponent estimate is a finite-time effect.  To support this
assertion, we show simulation data for the dependence of $T_N$ versus $N$ in
Fig.~\ref{avm-T} for representative parameter values: (a) initial width of
the fitness distribution $F_0=1$ (b) $F_0=N$ and (c) $F_0=N^2$, and
$\delta\! f$, the change in individual fitness in a voting event, fixed to be
1.  The data in the figure are based on $10^4$ realizations for $N$ up to
$2^{14}=16384$.  On a double logarithmic scale, the data of $T_N$ versus $N$
appears relatively straight, which suggests that a linear fit is warranted.
However, there is a small but consistent downward curvature in the data, a
feature that becomes apparent by studying local slopes of $T_N$ versus $N$
based on $k$ successive data points (insets to Fig.~\ref{avm-T}).  The choice
of $k$ is important: for too-small $k$ values, successive local slopes
fluctuate strongly and cannot be reliably extrapolated, while for $k$ too
large, the systematic trend in the local slope is averaged away.  We find
that for $k=10$, there is a good compromise between minimizing statistical
fluctuations and uncovering systematic local trends.

In Figs.~\ref{avm-T}(a) \& (b), corresponding to $F_0=1$ and $F_0=N$
respectively, the local slope is non-monotonic in $N$.  The source of this
crossover behavior appears to be the broadening of the fitness distribution
as a function of time.  This leads to rank-changing events becoming
progressively less frequent.  When rank changes stop occurring, the dynamics
should be the same as the FVM, for which $T_N\sim N$.  However, consensus
interrupts this gradual crossover.  Conversely, for $F_0=N^2$, the initial
fitness distribution is sufficiently broad that rank-changing events never
occur.  The dynamics thus coincides with that of the FVM, for which
$T_N\sim N$.  For this case, the simulation data for the local slope appears
to extrapolate to a value that is close to the expected value of 1
(Fig.~\ref{avm-T}(c)).

\begin{figure}[h]
\centerline{
	\subfigure[]{\includegraphics[width=0.33\textwidth]{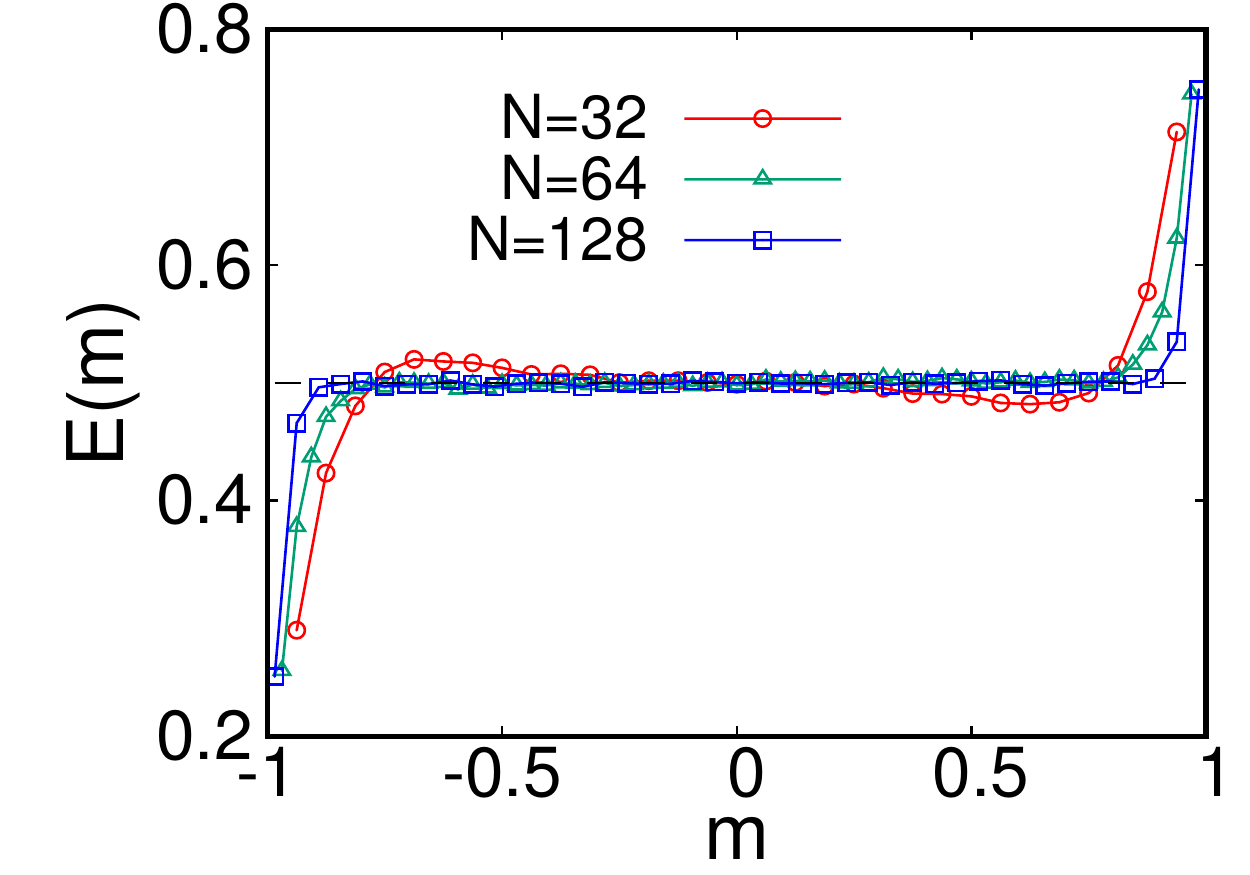}}
	\subfigure[]{\includegraphics[width=0.33\textwidth]{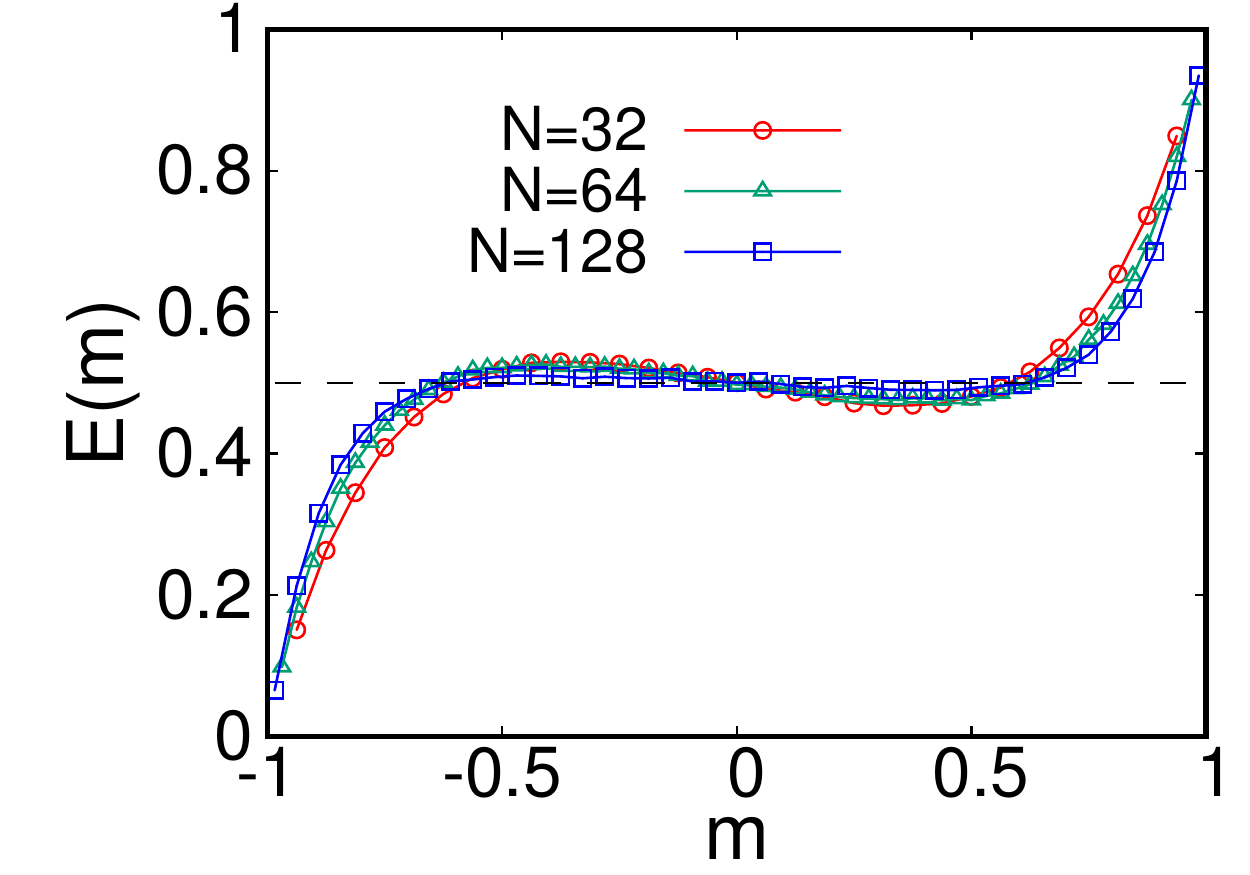}}       
	\subfigure[]{\includegraphics[width=0.33\textwidth]{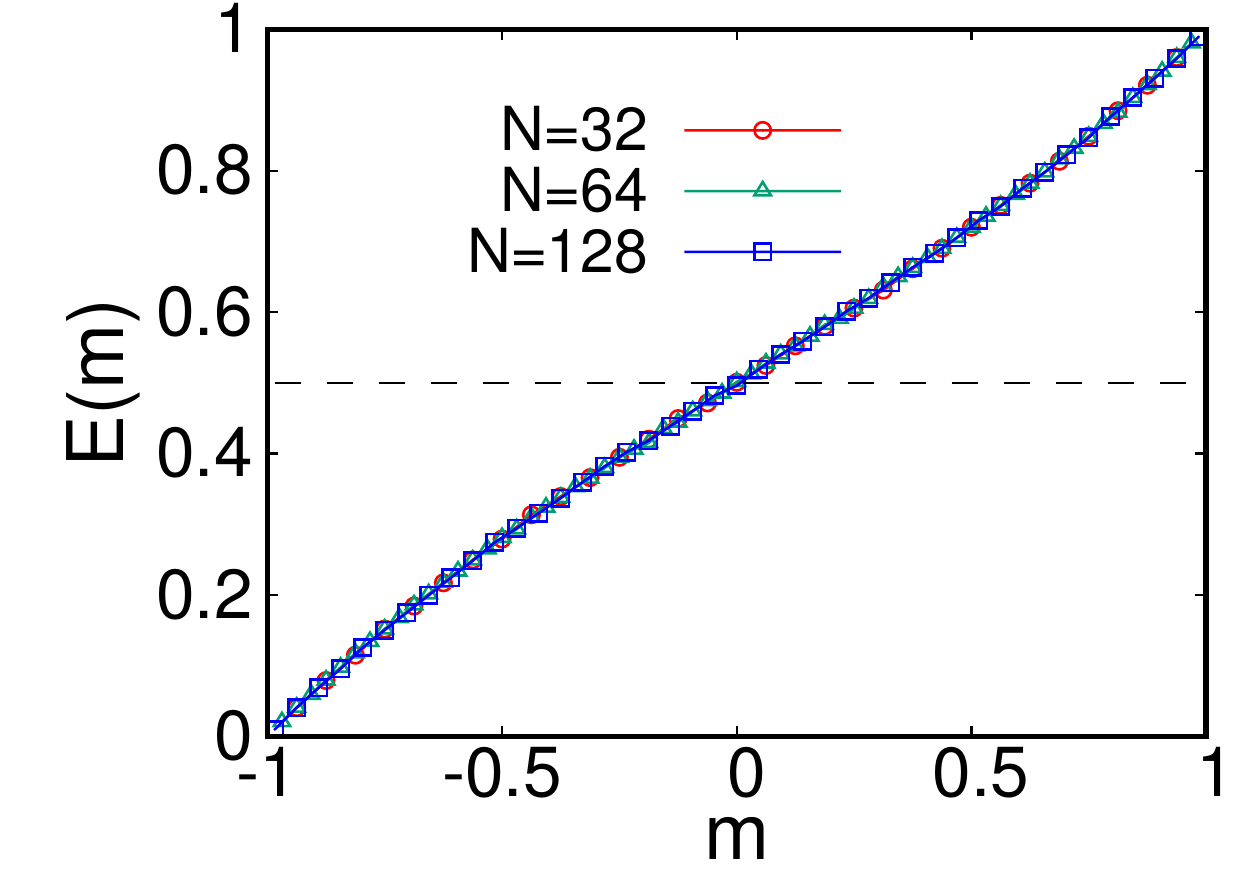}}
    }
    \caption{Exit probability as a function of initial magnetization $m$ for
      the AVM on the complete graph of $N$ sites with: (a) $F_0=1$, (b)
      $F_0=N$, and (c) $F_0=N^{2}$, for $\delta\! f=1$ in (a)--(c).  These data
      are obtained by averaging over $10^5$ trajectories.}
\label{avm-E}
\end{figure}

Simulation results for the exit probability is shown in Fig.~\ref{avm-E} for:
(a) $F_0=1$, (b) $F_0=N$, and (c) $F_0=N^2$, with $\delta f=1$ in all cases.
In (a) and (b), the exit probability $E(m)$ is a non-linear function of $m$,
which means that the magnetization is not conserved.  The non-linearity
indicates that there is an effective bias in the dynamics that tends to drive
a population with non-zero magnetization back to the zero-magnetization state
and thus forestalls consensus.  Note the curious feature, for which we have
no explanation, is that $E(m)$ is non-monotonic in $m$ for small $N$.  When
$F_0=N^2$, the exit probability is linear in $m$.  As discussed above,
rank-changing events no longer occur for $F_0=N^2$, so that the dynamics
should be the same as the VM.

To summarize, in spite of the simplicity of the AVM update rule, its basic
properties are surprisingly complex.  When the initial fitness distribution
is sufficiently broad or equivalently, the fitness increment $\delta\!f$ in a
single voting effect is sufficiently small, rank-changing events do not
occur, so that the dynamics is the same as the FVM, which, in turn, is the
same as that of the classic VM.  The dynamics of the AVM has a paradoxical
character in the time range where rank-changing events do occur.
Figure~\ref{avm-E} shows that the average magnetization is not conserved
because $E(m)$ strongly deviates from the form $E(m)=\frac{1}{2}(1+m)$ that
arises in the magnetization-conserving VM.  The non-linear dependence of
$E(m)$ in this figure indicates the presence of an underlying bias that tends
to drive the system to zero magnetization whenever $m\ne 0$.  In other
examples of voter-like models with non-conserved
magnetization~\cite{Lambiotte,Lambiotte3}, a similar non-linearity for $E(m)$
was observed.  As a result of the effective bias that drives the system to
zero magnetization, the consensus times in these models were found to grow
faster than a power law in $N$~\cite{Lambiotte,Lambiotte3}.  The observation
of an apparent power-law dependence of $T_N$ on $N$ found above and in
Ref.~\cite{Woolcock} is possibly a manifestation of the gradually diminishing
effective bias.  The main message from our analysis is that the exponent
$\alpha$ in $T_N\sim N^\alpha$ is strongly $N$-dependent and less than the
value 1.45 reported in~\cite{Woolcock}.

\subsection{Dynamical Non-Stationarity}

By directly adapting the theory given in~\cite{ben2005dynamics,Ben-Naim2006}
for the fitness distribution in a model of social competition, the
distribution of individual fitnesses in the AVM approaches a uniform
distribution in $[0,F(t)]$, with $F(t)=F_0+\,\delta\! f\, t/2$.  Consequently,
as the fitness distribution broadens, changes in fitness rank become more
rare.  When rank changes can no longer occur, the subsequent dynamics
approaches that of the FVM.

To understand this transition, we estimate the time dependence of
fitness-rank changes.  Consider two voters $i$ and $j$ of adjacent ranks,
with $f_i(0)>f_j(0)$; that is, voter $i$ is initially fitter than voter $j$.
Their fitnesses $f_i$ and $f_j$ at a later time $t$ are
\begin{align}
\label{fij}
\begin{split}
  f_i(t) &=f_i(0)+v_it \pm \sqrt{Dt}\,,\\
  f_j(t) &=f_j(0)+v_jt \pm \sqrt{Dt}\,.
\end{split}
\end{align}
Here $v_i$ is the systematic change in fitness because a higher-ranked voter
typically is more influential than a lower-ranked voter.  The ``speed'' $v_i$
at which the $i^{\rm th}$ voter gains fitness is proportional to the fraction
of voters with lower fitness.  For a uniform fitness distribution,
$v_i = f_i\, \delta\! f/F$.  Thus the speed of the best-ranked voter is
$\delta\! f$ and that of the worst-ranked voter is 0.  The term
$\pm\sqrt{Dt}$ denotes the change in fitness due to stochastic effects, which
give rise to rank-changing events.

\begin{figure}[ht]
\centerline{\includegraphics[width=0.45\textwidth]{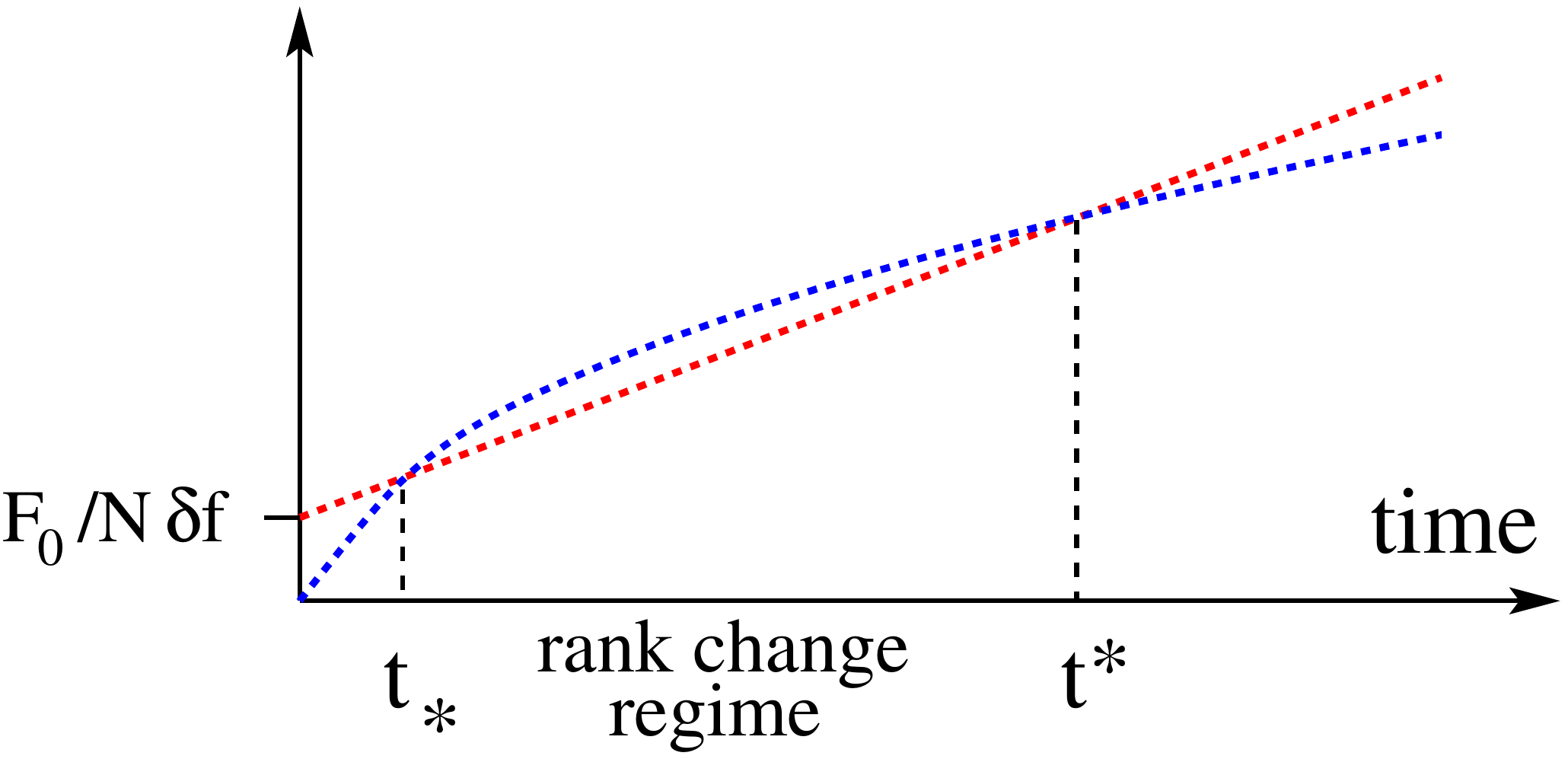}}
\caption{Schematic of the left-hand and right-hand sides of
  Eq.~\eqref{criterion} (red and blue respectively).  For this example, rank
  changes can occur only in the intermediate time regime between $t_*$ and
  $t^*$.}
\label{cartoon}
\end{figure}

In the absence of stochasticity, no rank-changing events occur.  To assess
the role of stochasticity on rank changes, we assume a negative stochastic
term for $f_i$ and a positive stochastic term for $f_j$ and find the
condition under which the ranks of these two voters can
switch~\cite{TW83,KR84}.  That is, suppose that at some time $t$,
$f_i(t)<f_j(t)$.  From Eq.~\eqref{fij}, this criterion gives
\begin{subequations}
  \begin{align}
    \label{ci}
  f_i(0)-f_j(0)+(v_i-v_j)t < \sqrt{4Dt}\,.
\end{align}
Now $v_i-v_j=\delta\!f/N$, while the diffusion coefficient associated with
the stochasticity is proportional to $(\delta\! f)^2$.  Thus
Eq.~\eqref{ci} becomes
\begin{align}
  \label{criterion}
  \frac{F_0}{N}+\frac{\delta\! f}{N} t < \delta\! f \,\sqrt{4t}\,.
\end{align}
\end{subequations}
Dividing through by $\delta\! f$, defining $a={1}/{N}$, and
$b=F_0/(N \delta\! f)$, the solution to \eqref{criterion} is
\begin{align}
  t=\frac{1}{2a^2}\big[(1-2ab)\pm\sqrt{1-4ab}\big]\,.
\end{align}

There are no solutions for $4ab>1$, which translates to $F_0/\delta\!f>N^2$.
That is, for a given $N$, if either the initial fitness range is sufficiently
large or the fitness change in a single voting event is sufficiently small,
no rank changes occur.  In this limit, the dynamics of the AVM reduces to the
FVM, which, in turn, is the same as the VM.  For $4ab<1$, the physically
relevant situation is $4ab\ll 1$.  Now there are two solutions:
\begin{align}
  t_* \approx \left(\frac{F_0}{N\delta\!f}\right)^2\,,\qquad\qquad t^*\approx
  N^2\,.
\end{align}
Between these two times, rank-changing events occur.  We may estimate the
time dependence of the number of rank changes as follows.  The typical
fitness difference of neighboring-ranked voters at time $t$ is
$\Delta\equiv F(t)/N$.  In a single voting event, the typical number of rank
changes is $dr\approx {\delta\! f}/{\Delta}= N\delta\! f/F(t)$, as long as
$\delta\!f>\Delta$.  Thus we estimate the number of rank changes per unit
time as
\begin{subequations}
\begin{align}
  \label{rankchange}
  N_r(t)  \simeq \frac{N\delta\! f/F(t)}{\delta t}=\frac{N\delta\!f/\delta t}{F_0 +  \,\delta\! f\,t/2}\,.
\end{align}

\begin{figure}[ht]
\centerline{\includegraphics[width=0.5\textwidth]{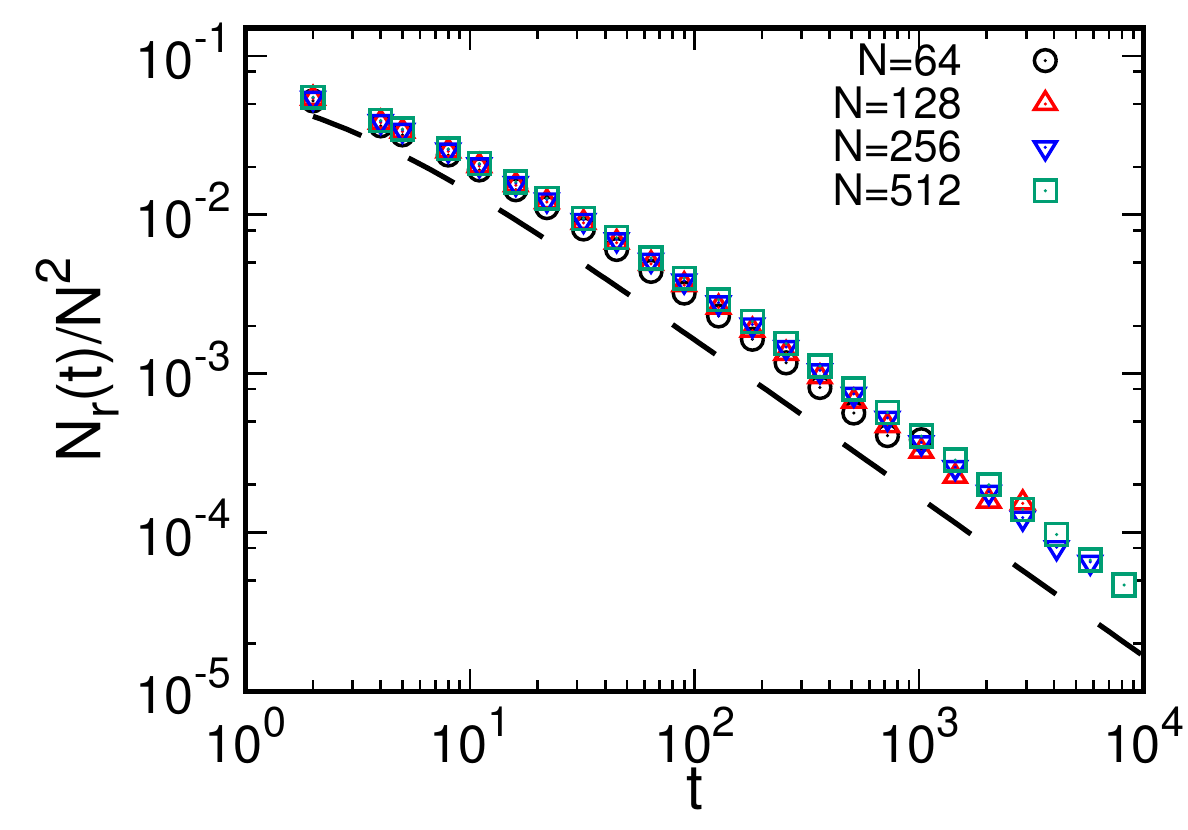}}
\caption{Time dependence of the number of rank changes per unit time,
  $N_r(t)$, scaled by $N^2$ in the AVM for $F_0=1$ and $\delta f =1$.  Data
  for $t>T_N$ are dominated by noise and are not shown.  The data are
  generated by averaging over $10^5$ realizations.  The dashed line is the
  prediction in Eq.~\eqref{rankchange2}.}
 \label{avm-rank}
\end{figure}

We can make this estimate more precise by computing the number of rank
changes averaged over the uniform distribution of fitnesses.  Consider the
case where $F_0=1$ and $\delta\! f =1$.  For the first voting event between
two voters with fitnesses $f_i$ and $f_j<f_i$, their fitnesses after the
voting event will be $f_i+1$ and $f_j$ respectively.  The number of rank
changes due to these changes is $dr=N(1-f_i)$.  Averaging this expression
over the uniform distribution of fitnesses subject to the constraint $f_i>f_j$,
gives $dr=N/3$.  Then using $\delta t= 4/N$ as the time increment for this
first voting event, the initial number of rank changes per unit time is
$N^2/12$.  Using this for $N_r(t\!=\!0)$ in \eqref{rankchange}, the number of
rank changes per unit time at any later time is
\begin{align}
  \label{rankchange2}
  N_r(t)  =\frac{N^2\delta\! f/12}{F_0 +  \,\delta\! f\,t/2}\,.
\end{align}
\end{subequations}
This prediction is consistent with the simulations shown in
Fig.~\ref{avm-rank}.

The simple reasoning given above shows that the dynamics of the AVM is non
stationary.  At early times, rank-changing events occur frequently (as long
as $\delta\! f$ is not pathologically small) and these rank changes are
responsible for the slow approach to consensus.  However, at sufficiently
long times, rank changing events stop occurring and the dynamics crosses over
to that of the FVM.  Thus over a substantial time range the dynamics of the
AVM is governed by crossover effects.

\subsection{Magnetization Zero Crossings}

\begin{figure}[ht]
\centerline{\includegraphics[width=0.5\textwidth]{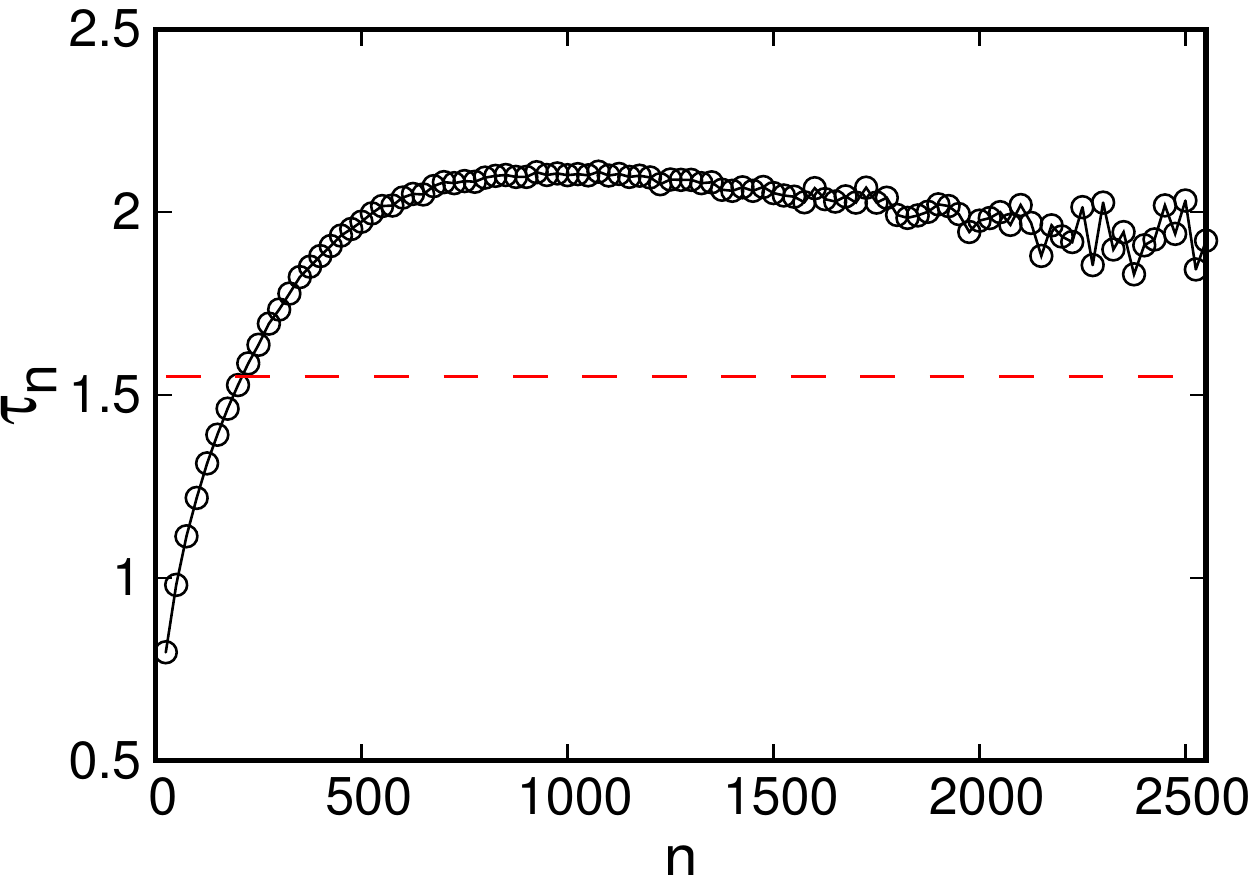}}
\caption{Dependence of $\tau_n$, the $n^{\rm th}$ zero-crossing time on $n$
  for $10^6$ realizations with $N=256$.  The data are smoothed by averaging
  over 15 successive points.  The parameters are $F_0=1$ and $\delta\! f=1$.
  The dashed line is the exact zero-crossing time for the VM (\ref{zct-vm}).
  The average number of zero crossings is 900.3. }
 \label{avm-zero}
\end{figure}

The non-stationarity of the AVM also manifests itself in the times between
successive zero crossings of the magnetization.  For a system that starts at
zero initial magnetization, there are typically multiple instances when the
magnetization returns to zero before consensus is reached.  We define
$\tau_n$ as the average time between the $(n-1)^{\rm st}$ and $n^{\rm th}$
zero crossing, with the $0^{\rm th}$ crossing occurring at $t=0$.  Each
$\tau_n$ is averaged over those trajectories that have not yet reached
consensus by the $n^{\rm th}$ crossing.  A basic feature of these
magnetization zero crossings for the AVM is that $\tau_n$ varies
non-monotonically with $n$ (Fig.~\ref{avm-zero}).  In this plot, the number
of ``surviving'' trajectories decreases as $n$ increases (roughly a fraction
$10^{-3}$ of all realizations survive until $n=2500$), and the behavior of
$\tau_n$ becomes progressively noisier.  In contrast, the dynamics of the
classic VM is stationary and successive zero-crossing times are all the same;
the derivation of the crossing time for the VM is given in \ref{zct-vm}.

We can qualitatively understand the non-monotonicity of the AVM zero-crossing
times in terms of the time dependence of the rank changes of the voters.  As
derived in Eq.~\eqref{rankchange2}, rank changes are frequent at early times
and become progressively less common.  These rank changes give rise to an
effective bias $v(m)$ towards zero magnetization (see also the next section).
At early times, these frequent rank changes imply a strong bias to zero
magnetization; this leads to zero-crossing times that are smaller than in the
VM.  At later times, we can assess the role of the bias on magnetization
trajectories in terms of the P\'eclet number~\cite{P03},
$P\!e\equiv |v(m)m|/D(m)$, where $D(m)$ is the diffusion coefficient
associated with the trajectories.  As time increases and the bias becomes
weaker, only those trajectories that approach close to $m=\pm 1$ experience a
P\'eclet number $P\!e>1$ and get driven back towards zero magnetization.
These large-deviation trajectories lead to a zero-crossing time that is
larger than that of the VM.  Finally, at late times ($t \gg N^2$), rank
changes become sufficiently rare that the dynamics approaches that of the VM
and the zero-crossing times also approach that of the VM.  This asymptotic
limit will be reached only when the number of zero crossings $n$ is of the
order of $N^2$ when rank changes no longer occur.

\section{Dynamics of  the Reputational Voter Model (RVM)}
\label{RVM}

\subsection{Effective Potential}

In RVM, each voting event leads to a fixed number of rank changes
(Fig.~\ref{rvm}).  This implies that the dynamics is stationary, which
simplifies the analysis of this model.  We will argue that the dynamics of
the magnetization is equivalent to that of a random walk that is confined to
an effective potential well, leading to an anomalously long consensus time
compared to the VM and the AVM.

In a single voting event, the magnetization $m$ changes by
$\delta m\equiv \pm 2/N$ and the average time for such a voting event is
$\delta t=N/(N_+N_-)$, where $N_\pm$
are the number of voters in the $+$ and $-$ states, respectively.  We define
$w(m\to m')$ as the probability that the magnetization changes from $m$ to
$m'$ in a single voting event and $P(m,t)\delta m$ as the probability that
the system has a magnetization between $m$ and $m+\delta m$.  The
Chapman-Kolmogorov equation for the time dependence of $P(m,t)$ is
\begin{subequations}
\begin{align}
P(m,t\!+\!\delta t)\!=\! w(m\!-\!\delta m\to m)P(m\!-\!\delta m,t) + w(m\!+\!\delta m\to m)P(m\!+\!\delta m,t).
\end{align}
Expanding this equation to second order in a Taylor series gives the
Fokker-Planck equation
\begin{align}
 \frac{\partial}{\partial t} P(m,t)=-\frac{\partial }{\partial m}[v(m)P] + \frac{\partial^2 }{\partial m^2}[D(m)P]\,,
 \label{FP}
\end{align}
\end{subequations}
where the drift velocity and diffusion coefficient are given by
\begin{align*}
  v(m)&=2[2w(m\!\to\! m+\delta m)-1]/(N\delta t)=[2w(m\to m+\delta m)-1](1-m^2)/2\,,\\
  D(m)&= 2/(N^2 \delta t)=(1-m^2)/2N\,,
\end{align*}
and where the second equalities follow by expressing the time step
$\delta t =N/(N_+N_-)$ in terms of the magnetization,
$\delta t= 4/[N(1-m^2)]$.

\begin{figure}[ht]
\centerline{\includegraphics[width=0.6\textwidth]{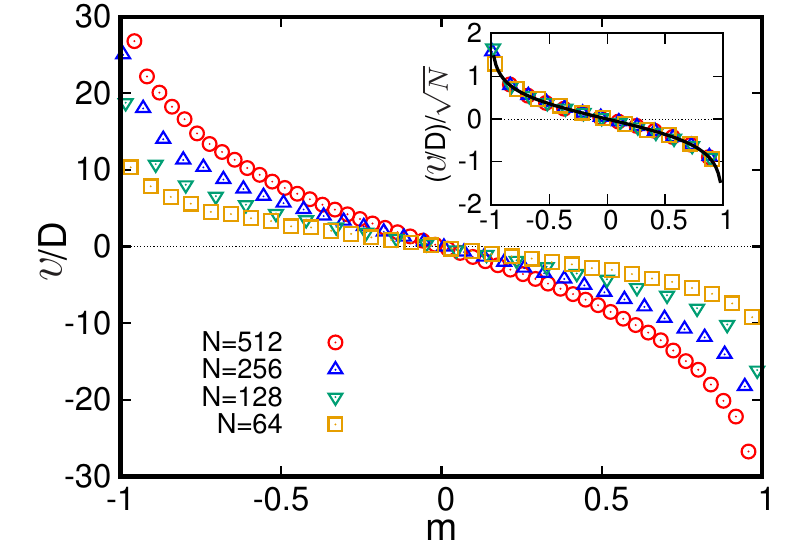}}
\caption{Dependence of $v(m)/D(m)$ for the RVM on $m$ for different $N$.  The
  inset shows the data collapse when $v/D$ is divided by $\sqrt{N}$.  The
  solid curve is the empirical fit $f(m)=-0.65\, \text{tanh}^{-1}\,m$ (see
  text).  The data represent averages over $10^4$ realizations.}
\label{rvm-bias}
\end{figure}

In Fig.~\ref{rvm-bias}, we plot the ratio $v(m)/D(m)$ versus $m$.  For this
data, we take the initial magnetization to be zero, and define the initial
average ranks of voters in the $+$ and $-$ states to be equal.  The quantity
$w(m \to m+\delta m)$ is measured as the probability that the magnetization
of the system increases from $m$ to $m+\delta m$.  The important feature is
the non-zero drift velocity that drives the population \emph{away} from
consensus and ultimately leads to a long consensus time.  Empirically, we
also find that the curves of $v/D$ for different $N$ all collapse onto a
single universal curve when the data is scales by $\sqrt{N}$ (inset to
Fig.~\ref{rvm-bias}).  The resulting scaled curve has a sigmoidal shape that
is turned on its side.  We find, therefore, that this curve is well fit by
the archetypal sigmoidal function $f(m)= - 0.65\,\text{tanh}^{-1}\,m$, where
the amplitude 0.65 gives the minimum deviation between the data for $v/m$ and
the fit.

\subsection{Consensus Time and Exit Probability}

\begin{figure}[ht]
  \centerline{\subfigure[]{\includegraphics[width=0.475\textwidth]{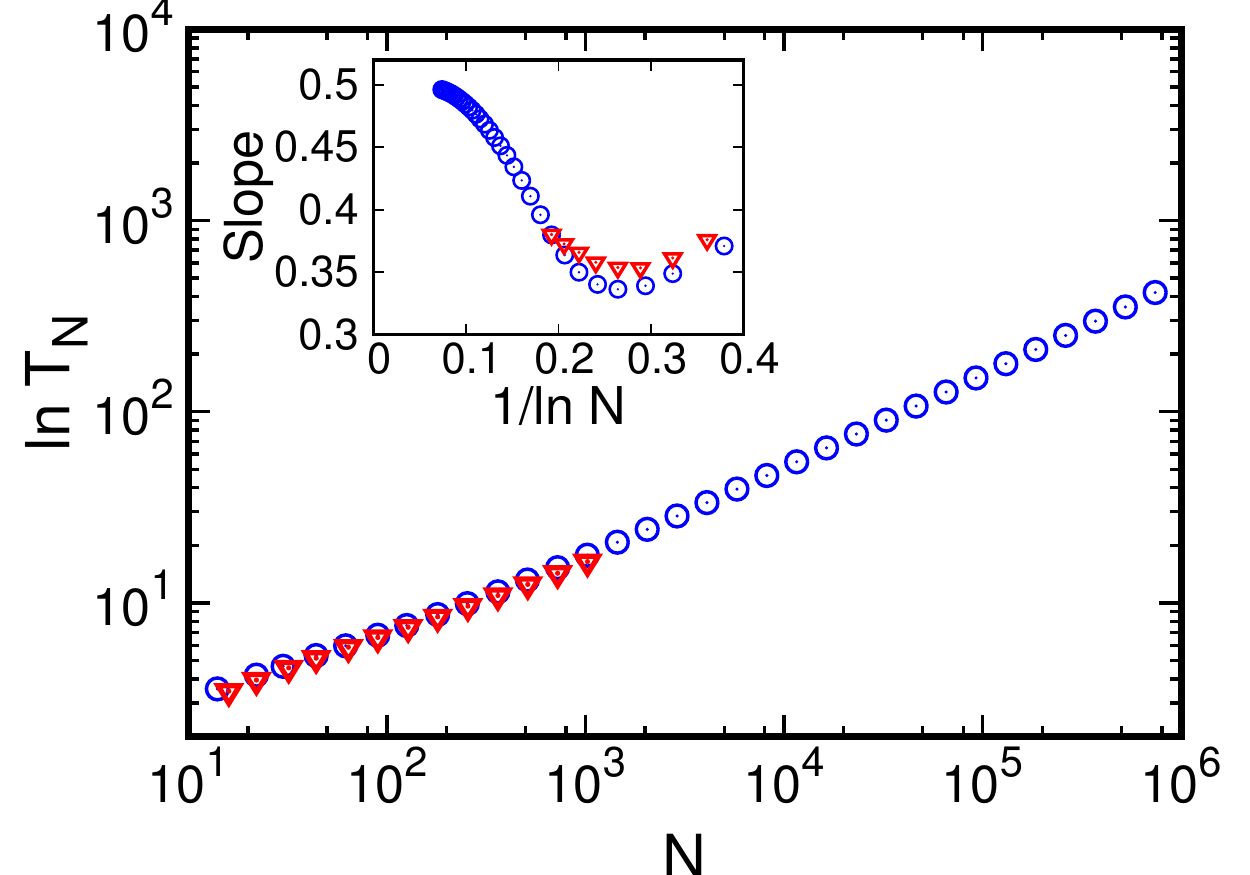}}\quad
    \subfigure[]{\includegraphics[width=0.475\textwidth]{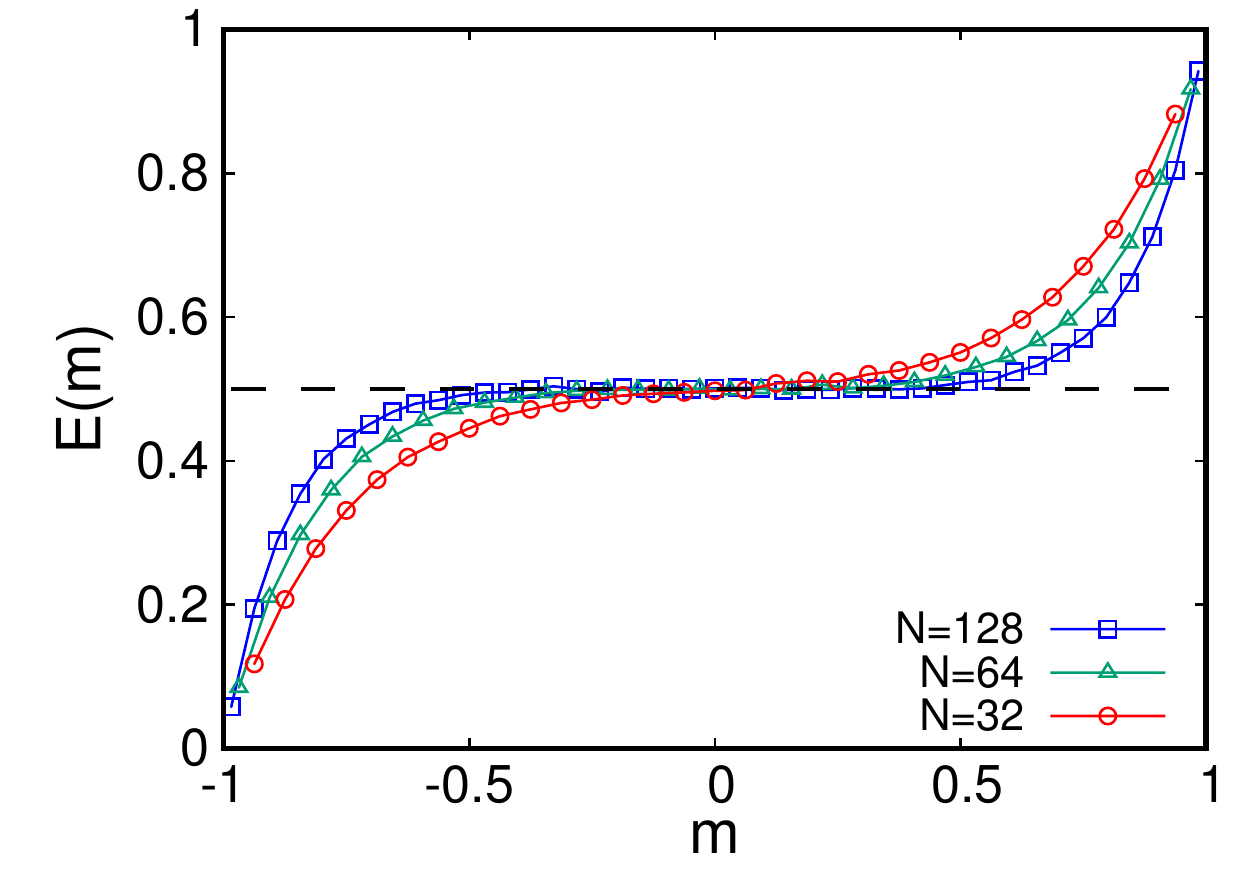}}}
  \caption{(a) Dependence of $\ln T_N$ versus $N$ on a double logarithmic
    scale based on: (i) $10^4$ realizations of the RVM (red triangles), and
    (ii) numerical integration of Eq.~\eqref{mfpt-rvm} (blue circles).  The
    inset shows the local slopes of these two datasets as a function of
    $1/\ln N$. (b) Exit probability of the RVM as a function of initial
    magnetization $m$ for different $N$. These data are based on $10^5$
    realizations. }
  \label{rvm-fpt} 
\end{figure}

Because the drift velocity drives the system away from consensus, we
anticipate that the consensus time will grow faster than a power law in $N$,
as shown in Fig.~\ref{rvm-fpt}.  For this data, the initial magnetization is
set to $m=0$ and the voter ranks are chosen so that the average ranks of $+$
and $-$ voters are, on average, equal.  The data in this figure indicate that
$T_N$ grows faster than a power law in $N$.  There is also an extremely slow
crossover to the asymptotic behavior (inset to Fig.~\ref{rvm-fpt}(a)) and it
is not possible to determine the functional form of $T_N$ based on simulation
data up to $N=1024$.

To give a more principled and reliable estimate for the $N$ dependence of
$T_N$, we write the backward Kolmogorov equation for the consensus
time~\cite{R01,Gardiner}
\begin{subequations}
\begin{align}
  T_N(m)=w(m\to m+\delta m)T_N(m+\delta m)+w(m\to m-\delta m)T_N(m-\delta m)+\delta t\,.
\end{align}
In the continuum limit this recursion becomes \cite{R01,Gardiner}
\begin{align}
  \label{T-RVM}
  \frac{v(m)}{D(m)}\frac{\partial T_N(m)}{\partial m} +
  \frac{\partial^2 T_N(m)}{\partial m^2}=-\frac{1}{D(m)}\,.
\end{align}
\end{subequations}
For arbitrary functional forms of $v(m)$ and $D(m)$, the formal solution of
\eqref{T-RVM} is~\cite{Gardiner}
\begin{align}
  \label{mfpt-rvm}
  T_N(m)=\int^{1}_m e^{-{A}(m')} \left[\int^{m'}_0
  \frac{e^{{A}(m'')}}{D(m'')}dm''\right]dm'\,,
\end{align}
where ${A}(m)=\int^{m}_0 [v(m')/D(m')] dm'$.  While it is generally not
possible to solve \eqref{mfpt-rvm} analytically, we can numerically integrate
this equation.  Here, we use our empirical observation that
$v(m)/D(m) = \sqrt{N} f(m)$, with $f(m)= -0.65\,\text{tanh}^{-1}\,m$ (inset
to Fig.~\ref{rvm-bias}).  The outcome of this numerical integration for $N$
up to $10^6$ is also shown in Fig.~\ref{rvm-fpt}(a).  The simulation data and
the integration data are nearly the same, and the local slopes of these two
datasets show similar behaviors.  However, since we can obtain integration
data up to $N=10^6$, we can now see the asymptotic trend in the local slope,
which indicates that the local slope eventually converges to $\frac{1}{2}$
(inset to this figure).  Thus we argue that the consensus time for the RVM
has the dependence $T_N\sim \exp(\sqrt{N})$.

Due to the non-zero drift velocity in the RVM, the magnetization is not
conserved, a feature that again manifests itself in the non-linear dependence
of the exit probability $E(m)$ on initial magnetization
(Fig.~\ref{rvm-fpt}(b)).  We again define the initial state so that the ranks
of $+$ and $-$ voters are equal, on average.  As $N$ increases, $E(m)$
gradually approaches a step function; this is a consequence of $v(m)/D(m)$
being an increasing function of $N$.  The step-like form of $E(m)$ is also
consistent with the consensus time growing faster than any power law in $N$
as shown in Fig.~\ref{rvm-fpt}(a).

\section{Summary and Discussion}
\label{conclude}

We studied a set of voter-like models on the complete graph, corresponding to
the mean-field limit, in which each voter has a characteristic fitness that
is a measure of its influence on others.  Our motivation for investigating
these models is that, in real life, some individuals are influential and
other less so; moreover, the influence of each individual can change with
time as opinions evolve.  While our models are highly stylized, perhaps they
provide a useful first step to understand the role of individual
persuasiveness on how opinions change in a population.

For the fitness voter model (FVM), where the fitness of each voter is
distinct and fixed, a simple, but striking result is that its
voting dynamics turns out to be identical to the classic VM.  Our main focus
was on voter models in which the fitness of each voter, as well as its
opinion, can change in an elemental update event.  We found that the coupled
dynamics of the fitness and voting state of each voter leads to rich dynamics
and also to very slow and subtle crossover effects.  This type of coupled
dynamics between voting state and fitness also shares some conceptual
commonality with voter models in which the connections between voters can
change as their opinions in each
update~\cite{gross2006epidemic,holme2006nonequilibrium,kozma2008consensus,shaw2008fluctuating,durrett2012graph,rogers2013consensus,Woolcock}.

We investigated two examples in which changing individual voter fitness
controls the consensus dynamics.  In the adaptive voter model (AVM), the
fitness of the influencer voter increases by a fixed amount while the
fitness of the influenced voter is unchanged in a single voting event.  This
same model was recently investigated in Ref.~\cite{Woolcock}, where it was
reported that the consensus time $T_N\sim N^\alpha$, with
$\alpha\approx 1.45$.  We argued instead that the dynamics of the AVM is more
subtle than this simple power law.  In particular, the dynamics has a
non-stationary character, in which the fitness distribution of the population
broadens in time.  Eventually, the width of this distribution broadens to the
point where fitness updates no longer change the relative ranks of individual
voters.  When this occurs, the opinion dynamics slowly crosses over to that
of the FVM, which in turn is the same as the classic VM.  This crossover is
interrupted by consensus, and the dependence of $T_N$ on $N$ appears to be a
power law, but with an exponent that is smaller than 1.45 (Fig.~\ref{avm-T}).

We also introduced the reputational voter model (RVM), which has the
advantage that its dynamics is stationary.  In the RVM, each voter is
assigned a unique integer-valued rank that ranges from 1, for the best-ranked
voter, to $N$, for the worst-ranked voter.  In an update, the influencer
voter moves up by 1 in rank while the rank of the influenced voter is
unchanged.  A  salient feature of this dynamics is that the voters in the
minority opinion tend to be higher ranked than those with the majority
opinion.

\begin{figure}[ht]
  \centerline{\includegraphics[width=0.475\textwidth]{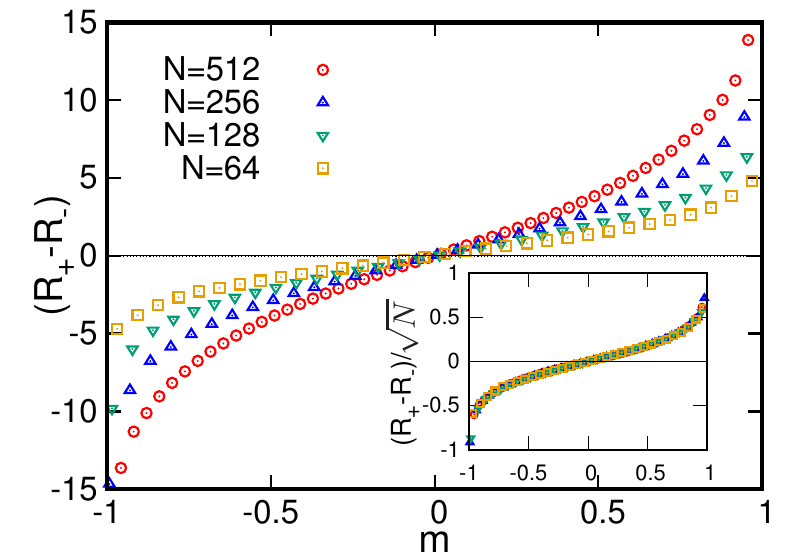}} 
  \caption{Difference between the average rank of $+$
    and $-$ voters, $R_+-R_-$, as a function of $m$ for different $N$.  The
    inset shows the data collapse when $R_+-R_-$ is scaled by $\sqrt{N}$.
    The data represent averages over $10^4$ realizations.}
  \label{rvm-ep}
\end{figure}

Consider a single update event, in which a voter with a $-$ opinion is
converted to $+$.  For this to occur, the reputation of this $-$ voter must be
lower than the $+$ voter.  After this update, the average rank of the $+$
voters becomes a bit worse: the influencer voter moves up by one rank, but
the influenced voter, whose rank is typically much lower, now joins the list
of $+$ voters.  Concomitantly, the $-$ voters have lost one voter whose
average rank is low, so that the average rank of this group improves.  We
have not been able to go beyond this heuristic observation to compute the 
magnitude of the rank difference as a function of the magnetization.
Nevertheless, the trend from Fig.~\ref{rvm-ep} is clear: for nonzero $m$, the
minority voters are better ranked and for fixed $m$ this rank difference
appears to grow as $\sqrt{N}$ (inset to Fig.~\ref{rvm-ep}).  This rank
difference is the mechanism underlying the drift velocity that drives the
system away from consensus.  The primary consequence of this bias is that
$T_N$ grows faster than a power law in $N$ and the numerical evidence
suggests that $T_N\sim\exp(\sqrt{N})$ (Fig.~\ref{rvm-fpt}(a)).

There are multiple ways in which fitness, or rank changes of voters can be
implemented; we only treated the case where the influencer voter becomes
``stronger'', while the influenced voter is not affected.  It is also natural
to consider the cases where: (i) the influencer voter becomes stronger and
the influenced voter becomes weaker, and (ii) the influencer voter is
unaffected and only the influenced voter becomes weaker.  In case (i),
simulations indicate that the dynamical behavior is similar to the situation
where only the influencer voter becomes stronger.  In case (ii), however,
the dynamics appears to be in the same universality class as the VM.  Namely,
the consensus time $T_N\sim N$ and the exit probability $E(m)=(1+m)/2$.  The
latter behavior arises because the highest-ranked voter does not change its
opinion throughout the dynamics, a situation that also arises in the FVM.

\section*{Acknowledgments} 

We gratefully acknowledge financial support from NSF grant DMR-1608211.

\appendix
\section{Equivalence between the FVM and the VM}
\label{phi}

For notational simplicity, let $L$ and $M=N-L$ denote the number of voters
with $+$ and $-$ opinions, respectively. We define the probability for $L$ to
increase by 1 in a time step $\delta t$ as $\phi(L\rightarrow L+1)$ and
$\phi(L\rightarrow L-1)=1-\phi(L\rightarrow L+1)$ for the probability for $L$
to decrease by 1.  We also define $f$ and $g$ to represent the fitness of
voters with $+$ and $-$ opinions, respectively.  We define a voter
configuration as
\begin{align}
  C(\mathbf{f},\mathbf{g},L+1) \equiv   \{(f_1,f_2, \ldots f_{L+1})(g_1,g_2, \ldots g_{M-1})\}
  \label{conf}
\end{align}
in which there are $L+1$ voters with $+$ opinion (left set in $C$) and $M-1$
voters with $-$ opinion (right set).  In each set, we order the individual
fitnesses so that $f_i<f_{i+1}$ and $g_i<g_{i+1}$.  When the number of $+$
voters increases from $L$ to $L+1$, the system moves from one of
$ \mathcal{N}(L)=\binom{N}{L}$ configurations to one of
$\mathcal{N}(L+1)=\binom{N}{L+1}$ configurations.

We focus on one such event in which $L$ increases and thereby system ends in
the configuration $C(\mathbf{f},\mathbf{g},L+1)$ specified in
Eq.~\eqref{conf}. In this event, one out the $L+1$ voters which are currently
in $+$ opinion set of $C(\mathbf{f}, \mathbf{g},L+1)$, must have left $-$
opinion set of the previous configuration.  Let $f_i$ ($1\leq i \leq L+1$) be
fitness of this relocated voter and denote $C_i(\mathbf{f}',\mathbf{g}',L)$
as the previous configuration from which system reached
$C(\mathbf{f}, \mathbf{g},L+1)$. It is important to note: (a) the difference
between $C(\mathbf{f},\mathbf{g},L+1)$ and $C_i(\mathbf{f}',\mathbf{g}',L)$
is only the opinion of the voter with fitness $f_i$; (b) in both
configurations, the number of $+$ opinion voters whose fitness is larger than
$f_i$ is $L+1-i$.  Using these two facts, the probability for the system to
move from $C_i(\mathbf{f}',\mathbf{g}',L)$ to $C(\mathbf{f},\mathbf{g},L+1)$
is given by
\begin{align}
  \label{hi}
h_i=\frac{1}{\mathcal{N}(L)}\times \frac{1}{M} \times \frac{L+1-i}{L}\,. 
\end{align}
The first factor is the probability for the system to be in one out of
$ \mathcal{N}(L)$ possible configurations.  The second factor is the
probability that the voter with fitness $f_i$ is picked from $M=N-L$ elements
in the group of $-$ voters.  The third factor is the probability to pick a
voter in the $+$ set in $C_i(\mathbf{f}',\mathbf{g}',L)$ whose fitness larger
than $f_i$.

This probability $h_i$ should be summed over all $1\leq i \leq L+1$ to
enumerate all possibilities that lead to $C(\mathbf{f},\mathbf{g},L+1)$.
Thus the probability to reach $C(\mathbf{f},\mathbf{g},L+1)$ from all
eligible configurations by an event in which $L$ increases is
\begin{align}
 \label{eq3}
 \sum^{L+1}_{i=1}h_i = \frac{1}{\mathcal{N}(L)}\frac{L+1}{2M} \,.
\end{align}
This probability is independent of the fitness of the voters in
$C(\mathbf{f},\mathbf{g},L+1)$.  Therefore, in an event in which $L$
increases by 1, any one out of $\mathcal{N}(L+1)$ configurations can be
reached with the probability given in Eq.~\eqref{eq3}.

In Eq.~\eqref{hi}, we assumed that all configurations with $L$ voters in the
$+$ voting state have the same probability $1/\mathcal{N}(L)$.  This
assumption is justified because, at time $t=0$, all $\mathcal{N}(L=N/2)$
configurations are chosen with equal probability.  Because of
Eq.~\eqref{eq3}, at any later time all $\mathcal{N}(L)$ configurations with
fixed $L$ are visited by the system an equal number of times, on average.

In Eq.~\eqref{eq3} we found the probability to reach one out of
$\mathcal{N}(L+1)$ configurations by an event in which $L$ increases. To
obtain $\phi(L\rightarrow L+1)$ we must sum over all possibilities that
result in all $\mathcal{N}(L+1)$ different configurations.  Because each of
these probabilities are same (Eq.~\eqref{eq3}), we have
\begin{align}
\phi(L\rightarrow L+1)= \sum^{L}_{i=1} h_i \,\, \mathcal{N}(L+1)  =  \frac{1}{\mathcal{N}(L)}\frac{L+1}{2M}\mathcal{N}(L+1)=\frac{1}{2}.
\end{align}
That is, the probabilities for the number of $+$ opinion voters to increase
and decrease in a single time step are equal, i.e.,
$\phi(L\rightarrow L+1)=\phi(L\rightarrow L-1)=1/2$.  These are the same as
the transition probabilities in the VM, which establishes the equivalence
between the FVM and the classic VM.

\section{Zero crossings statistics of the magnetization in the VM}
\label{zct-vm}

For the VM, we want to determine: (a) the conditional time $T_-(m)$ for the
population to start at magnetization $m=0$ and return to $m=0$ without
reaching consensus, and (b) the conditional time $T_+(m)$ to start at $m=0$
and reach consensus without return.  These conditional exit times satisfy the
backward Kolmogorov equations
\begin{equation}
\label{de-Tpm}
D(m)\frac{d^2\big[E_\pm(m) T_{\pm}(m)\big]}{dm^2} = - E_{\pm}(m),
\end{equation}
subject to the boundary conditions: $E_\pm(0)T_{\pm}(0)=E_\pm(1)T_{\pm}(1)=0$.
Here $E_\pm(m)$ are the exit probabilities to $m=0$ and $m=1$ and are given
by $E_{+}(m)=m$ and $E_{-}(m)=1-m$.

The solutions to \eqref{de-Tpm} are
\begin{align}
\label{Tz-voter}
  E_+(m)T_+(m) &=  N \big[(1+m) \ln(1+m) - (1-m) \ln(1-m) - 2m \ln 2\big]\,, \nonumber \\
  E_-(m) T_-(m) &= 2N \big[2m \ln2-(1+m)\ln(1+m)\big] \,.
\end{align}

To obtain the escape and return times, we consider the initial condition
$m=2/N$, which is the outcome after a single voting event (where a $+-$ pair
changes to $++$) and include the time increment to go from $m=0$ to $m=2/N$.
Then from \eqref{Tz-voter}, the escape time is
\begin{subequations}
\begin{align}
\tau_e &= \frac{4}{N} + T_+\left(\frac{2}{N}\right)\nonumber \\
&=\frac{4}{N}+\frac{N^2}{2}\left[\left(1+\frac{2}{N}\right) 
\ln \left(1+\frac{2}{N}\right)- \left(1-\frac{2}{N}\right)
\ln \left(1-\frac{2}{N}\right) - \frac{4}{N} \ln 2 \right]\nonumber \\
&\approx  2N(1-\ln 2) +\mathcal{O}\left(\frac{1}{N}\right)\,.
\end{align}
Similarly, the return time, equivalent to the zero-crossing time, is
\begin{align}
\tau_0 &= \frac{4}{N} + T_-\left(\frac{2}{N}\right)\nonumber \\
&=\frac{4}{N}+\frac{2N}{1-\frac{2}{N}}\left[\frac{4}{N}\ln 2 -\left(1+\frac{2}{N}\right) 
\ln \left(1+\frac{2}{N}\right)\right]\nonumber \\
&\approx  4(2\ln 2-1) +\mathcal{O}\left(\frac{1}{N}\right).
\end{align}
\end{subequations}
Note that the average escape time $\tau_e$ is $\mathcal{O}\left(N\right)$,
while the average return time $\tau_0$ is $\mathcal{O}\left(1\right)$.

\newpage

\newcommand{\newblock}{}

\end{document}